\documentclass[aps,prl,amsmath,amssymb,superscriptaddress,showpacs,floatfix,preprint,showkeys]{revtex4-1}

\usepackage{graphicx}
\usepackage{amssymb}
\usepackage{amsmath}
\usepackage{dcolumn}
\usepackage{multirow}
\usepackage{bm}
\usepackage[latin1]{inputenc}
\usepackage[flushmargin,ragged]{footmisc}
\usepackage{rotating}
\usepackage{placeins}

\usepackage{xcolor,colortbl}

\begin{document}
\bibliographystyle{apsrev4-1}

\title{Group theory for structural analysis and lattice vibrations in phosphorene systems}

\author{J. Ribeiro-Soares}
\email[Author to whom correspondence should be addressed: ]{jenainassoares2@gmail.com }
\affiliation{Departamento de F\'{\i}sica, Universidade Federal de Minas Gerais, Belo Horizonte, MG, 30123-970, Brazil}
\affiliation{Departamento de F\'{\i}sica, Universidade Federal de Lavras, Lavras, MG, 37200-000, Brazil}

\author{R. M. Almeida}
\affiliation{Departamento de F\'{\i}sica, Universidade Federal de Minas Gerais, Belo Horizonte, MG, 30123-970, Brazil}

\author{L. G. Can\c{c}ado}
\affiliation{Departamento de F\'{\i}sica, Universidade Federal de Minas Gerais, Belo Horizonte, MG, 30123-970, Brazil}

\author{M. S. Dresselhaus}
\affiliation{Department of Electrical Engineering and Computer Science, Massachusetts Institute of Technology (MIT), Cambridge, MA 02139, USA and Department of Physics, Massachusetts Institute of Technology (MIT), Cambridge, MA 02139, USA}

\author{A. Jorio}
\affiliation{Departamento de F\'{\i}sica, Universidade Federal de Minas Gerais, Belo Horizonte, MG, 30123-970, Brazil}

\date{Submitted on \today}

\begin{abstract} Group theory analysis for two-dimensional elemental systems related to phosphorene is presented, including (i) graphene, silicene, germanene and stanene, (ii) dependence on the number of layers and (iii) two stacking arrangements. Departing from the most symmetric $D_{6h}^1$ graphene space group, the structures are found to have a group-subgroup relation, and analysis of the irreducible representations of their lattice vibrations makes it possible to distinguish between the different allotropes. The analysis can be used to study the effect of strain, to understand structural phase transitions, to characterize the number of layers, crystallographic orientation and nonlinear phenomena.
\end{abstract}



\maketitle

Bulk phosphorus allotropes have been studied for $100$ years \cite{bridgman1914two,warschauer1963electrical}, but it is only in the post-graphene era that their notable few-layer induced novel properties were placed under scrutiny. The two-dimensional (2$D$) monolayer counterpart of bulk black phosphorus \cite{haussermann2003high}, the phosphorene (including the recently proposed ``blue phosphorene'' \cite{Zhu2014semiconductingblue,Xie2014blueribbons}), and the few-layer related systems have generated intense theoretical and experimental efforts addressing their optical \cite{Zhu2014semiconductingblue,Xie2014blueribbons,Xia2014rediscoveringBP,Tran2014layercontrolledBGBP,Castellanos_Gomez2014IsolationBP,Rodin2014strain_induced_gap,Zhang2014ExtraordPLRamanBP,Pengke2014electrons}, mechanical \cite{Rodin2014strain_induced_gap,Fei2014LatVibRamanStrain,Lv2014ThermoelectricStrain}, thermal \cite{Fei2014ThermoelectricEffOrtho,Lv2014ThermoelectricStrain,Lv2014ThermoelectricPowerfactor} and electrical \cite{li2014black,Xia2014rediscoveringBP,liu2014phosphoreneUnexplored,Qiao2014highmobilityBP,koenig2014electricEFE,buscema2014fast,Engel2014BPphotodetector} properties. Black phosphorus (black P) has shown highly anisotropic properties \cite{Tran2014layercontrolledBGBP,Lv2014ThermoelectricPowerfactor,Lv2014ThermoelectricStrain,Fei2014ThermoelectricEffOrtho,Fei2014LatVibRamanStrain,Qiao2014highmobilityBP,Zhu2014semiconductingblue}, in addition to a band gap that increases with a decreasing number of layers [from $0.3$\,eV (bulk) to $1.45$\,eV for the monolayer \cite{liu2014phosphoreneUnexplored,Tran2014layercontrolledBGBP,Zhang2014ExtraordPLRamanBP}]. The role of strain in 2$D$ structures is another relevant aspect and it has been shown to be valuable to tune optical properties \cite{frank2010compression,conley2013bandgap,wang2013raman,Zhu2014semiconductingblue}.

Symmetry analysis has been used to understand the electronic structure of monolayer black P \cite{Pengke2014electrons}, and the Raman spectra of few-layer black P  \cite{Favron2014exfoliatingpristine}. Blue phosphorene (blue P), which was found to be compatible with the arsenic phase of phosphorus \cite{jamieson1963crystal,burdett1982pressure,haussermann2003high,Boulfelfel2012A17toA7phase}, is expected to exhibit a fundamental band gap that exceeds $2$ eV, while strain could be used to tune the band gap over a wide range of values \cite{Zhu2014semiconductingblue,Guan2014PhaseCoexistenceP}. A conversion route between the black and blue monolayer allotropes was already proposed, with stability and compatibility as in-plane heterostructures \cite{Zhu2014semiconductingblue,Guan2014PhaseCoexistenceP}.

In this letter, group theory is used to obtain symmetry-related information for 2$D$ black P, blue P, and related structures, including the dependence on the number of layers $N$ and on the stacking arrangements. The related structures are the other 2$D$ elemental structures, like graphene, and the Si, Ge and Sn analogs of graphene (silicene, germanene and stanene, respectively). Silicene and germanene have been recently synthesized \cite{Takeda1994SiGeanalogs,Vogt2012CompellingExSilicene,Li2014BuckledGermanene,Davila2014GermaneneAkin}, and stanene is theoretically predicted \cite{xu2013largegap}, and all these structures exhibit the same lattice structure as blue P monolayer. Therefore, the symmetry considerations for blue P discussed here are directly related to silicene, germanene, and stanene as well.

\begin{figure*}
        \centering
            \includegraphics[width=0.9\textwidth]{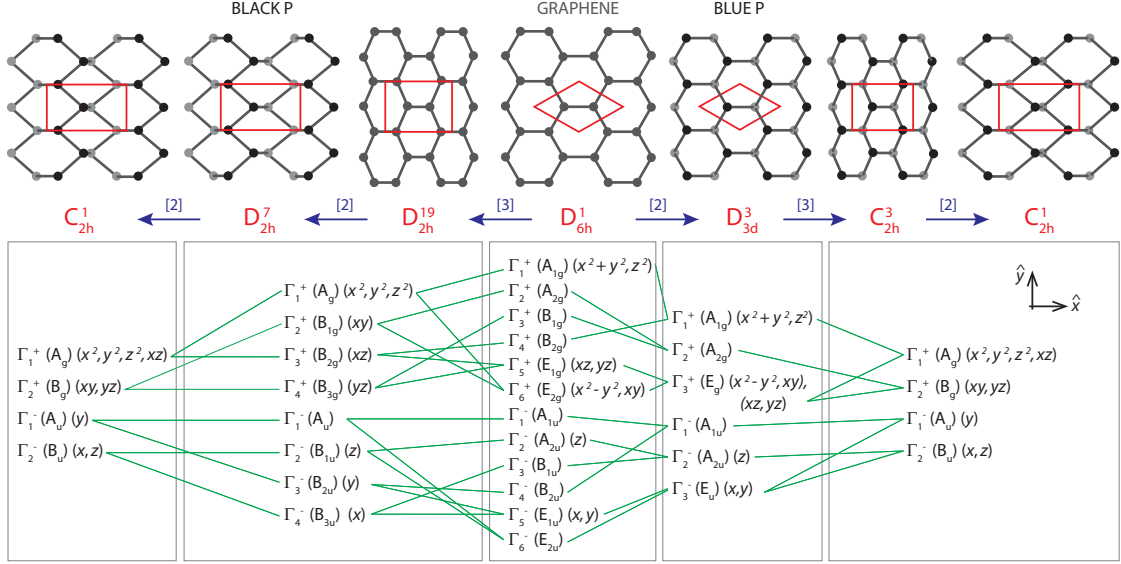}
        \caption{(Color online). Group-subgroup correlation for phosphorene-related systems. From center to left: correlation between graphene ($D_{6h}^{1}$ space group) and the black P monolayer ($D_{2h}^{7}$). The intermediate structure with the $D_{2h}^{19}$ space group (and the same $D_{2h}$ black P point group) occurs as an intermediate state between graphene and black P monolayer. A compressive strain in the $\hat{x}$ direction leads to a phase transition to the $C_{2h}^{1}$ subgroup. From center to right: correlation between graphene ($D_{6h}^{1}$ space group) and blue P monolayer ($D_{3d}^{3}$), followed by blue P monolayer with a compressive strain in the $\hat{x}$ direction ($C_{2h}^{3}$). In a further loss of centring translations, the $C_{2h}^{3}$ structure changes to the $C_{2h}^{1}$ group, with the same $C_{2h}$ point group. The top part brings structural schematics, while the bottom part displays the irreducible representations and the corresponding basis functions. The arrows indicate the sense of the lowering of symmetry, and the index of transformation of one group to another is indicated inside the brackets. To maintain the same axes orientation for the different structures, non-standard settings are used \cite{ITCA2005}. The character tables are given in \cite{[{See Supplementary Material at }][{ for character tables with the notation conversion from space group (SG) to point group (PG), the monolayer black P and blue P eigenvectors and a complete classification list of the modes as Raman active, infrared active, acoustic and silent, for all the structures and for $N$ layers.}]SupInfolinkGTPhosphorene}.
        }
        \label{correlationgraphene}
    \end{figure*}

Black and blue P monolayers present real space lattices that resemble the graphene honeycomb lattice, but in a puckered structure. Schematics for the blue and black P are given in the top part of Fig.~\ref{correlationgraphene}, and labeled accordingly. Light gray and black bullets indicate sets of atoms in different planes of the puckered structure. The red lines in the schematics show the top view of the unit cell. The black P monolayer primitive unit cell contains four atoms, while the blue P monolayer contains two atoms. The central schematics sketches the graphene structure, and dark gray bullets are used to indicate that all atoms are in the same plane.

Figure~\ref{correlationgraphene} shows a group-subgroup relation of the phosphorene parent materials, which are possible routes for structural modifications obtained via second-order phase transitions. Departing from the most symmetric $D_{6h}^1$ graphene space group, which appears at the center of the 2$D$ schematics in the top of Fig.~\ref{correlationgraphene}, the left-side route is started by an uniaxial compression, while the right-side route is started by an isotropic lattice compression.

From the schematics on the top of Fig.~\ref{correlationgraphene}, and departing from the $D_{6h}^1$ central graphene structure to the left route, uniaxial compression induces a phase transition to subgroup $D_{2h}^{19}$ of strained graphene. The hexagonal symmetry is lost, resulting in an orthorhombic structure where all atoms remain in the same plane. A possible and natural distortion of the orthorhombic lattice to accommodate uniaxial strain is the displacement of lines of atoms perpendicular to the strain direction, periodically up and down in the $\hat{z}$ direction (perpendicular to the 2D $\hat{x}\hat{y}$-plane structure), generating zigzag lines of atoms displaced to $+\hat{z}$  and $-\hat{z}$. Such a distortion generates the structure of black P, which belongs to the $D_{2h}^{7}$ subgroup. A final distortion along this route can be obtained by a shear force, which displaces the top and bottom set of atoms in opposite directions, breaking the bonding symmetry of atoms aligned along the strain direction and separating the zigzag lines of atoms that are displaced up and down. The system then undergoes a phase transition to the $C_{2h}^{1}$ subgroup.

Now on the right route shown in Fig.~\ref{correlationgraphene}, departing again from the $D_{6h}^1$ graphene structure, initially an isotropic strain changes the C-C bond distances, but it does not change the symmetry of the system (not shown in Fig.~\ref{correlationgraphene}). However, a possible and natural distortion of the hexagonal lattice to accommodate such a strain is the displacement of atoms, periodically up and down in the $\pm \hat{z}$ directions, generating a trigonal arrangement of atoms. Such a distortion generates the structure of blue P (also of silicene, germanene and stanene), which belongs to the $D_{3d}^{3}$ subgroup, with three fold rather than six fold rotational symmetry. The number of atoms in the unit cell is unchanged. In sequence, if at that stage an uniaxial strain is applied, the system loses the three-fold axis and undergoes a phase transition to the $C_{2h}^{3}$ orthorhombic subgroup for the strained blue P. A final hypothetical distortion on the $C_{2h}^{3}$ structure can lower the symmetry to a $C_{2h}^{1}$ subgroup (last schematics on the right in  Fig.~\ref{correlationgraphene}).

The structure for the last schematics on the right in Fig.~\ref{correlationgraphene} belongs to the same space group as the strained and sheared black P (last schematics on the left in Fig.~\ref{correlationgraphene}). Structural change between these two structures can be obtained via a first-order transition where the neighboring vertical lines of atoms exchange positions along $z$, possibly induced by a shear strain accompanied by compression perpendicular to the planes.


\begin{table}[h!]
\begin{center}
\caption{\label{tab:Gammavib} Irreducible representations for vibrational modes $\Gamma^{vib}$ in the phosphorene-related space groups.}
\begin{tabular}{cc}
\hline
\hline
$D_{2h}^{7}$ & $2\Gamma_{1}^{+} \oplus \Gamma_{2}^{+} \oplus 2\Gamma_{3}^{+} \oplus \Gamma_{4}^{+}\oplus \Gamma_{1}^{-} \oplus 2\Gamma_{2}^{-} \oplus \Gamma_{3}^{-} \oplus 2\Gamma_{4}^{-} $ \\
$D_{2h}^{19}$ & $\Gamma_{1}^{+}\oplus\Gamma_{2}^{+} \oplus \Gamma_{2}^{-} \oplus \Gamma_{3}^{+}\oplus \Gamma_{3}^{-} \oplus \Gamma_{4}^{-} $ \\
$D_{6h}^{1}$ & $\Gamma_{4}^{+}\oplus\Gamma_{6}^{+} \oplus \Gamma_{2}^{-} \oplus \Gamma_{5}^{-} $ \\
$D_{3d}^{3}$ & $\Gamma_{1}^{+}\oplus\Gamma_{3}^{+} \oplus \Gamma_{2}^{-} \oplus \Gamma_{3}^{-} $ \\
$C_{2h}^{3}$ & $2\Gamma_{1}^{+}\oplus \Gamma_{2}^{+} \oplus \Gamma_{1}^{-} \oplus 2\Gamma_{2}^{-} $ \\
$C_{2h}^{1}$ & $4\Gamma_{1}^{+}\oplus 2\Gamma_{2}^{+} \oplus 2\Gamma_{1}^{-} \oplus 4\Gamma_{2}^{-}$ \\
\hline
\hline
\end{tabular}
\end{center}
\end{table}

\begin{sidewaystable}%
\caption{\label{tab:GWVGamma} Space groups and irreducible representations for vibrational modes ($\Gamma^{vib}$), according to the allotrope (black P, blue P, silicene, germanene and stanene), to the number $N$ of layers and to the stacking order.}
\vspace{0.3cm}\tiny
\begin{ruledtabular}
\begin{tabular}{ccc}
\multicolumn{3}{c}{Black P}\\
\hline
   & $AA$ & $AB$   \\
\hline
$N$ odd &  &  [$D^{7}_{2h}$ ($Pbmn$, \#53)]~$ 2N(\Gamma^{+}_{1} \oplus \Gamma^{+}_{3} \oplus \Gamma^{-}_{2} \oplus \Gamma^{-}_{4}) \oplus N(\Gamma^{+}_{2} \oplus \Gamma^{+}_{4} \oplus \Gamma^{-}_{1} \oplus \Gamma^{-}_{3})$ \\
$N$ even & \multirow{-3}{*}{[$D^{7}_{2h}$ ($Pbmn$, \#53)]\footnote{\label{1}Notation: Schoenflies symbol, Hermann-Mauguin symbol, International Tables for Crystallography space group \#, Vol. A (ITCA) \cite{ITCA2005} - Vol. E (ITCE) \cite{ITCE2002} for``layered subperiodic groups'' could be used, but ITCA leads to an immediate comparison with the literature \cite{malard2009group,Ribeiro-Soares2014GTTMDCs}. One-to-one correlation exists when limited to the Brilluoin zone center.}~$ 2N(\Gamma^{+}_{1} \oplus \Gamma^{+}_{3} \oplus \Gamma^{-}_{2} \oplus \Gamma^{-}_{4}) \oplus N(\Gamma^{+}_{2} \oplus \Gamma^{+}_{4} \oplus \Gamma^{-}_{1} \oplus \Gamma^{-}_{3})$\footnote{\label{2}Space group (SG) notation. Conversion to point group (PG) notation, and convenient basis functions for each irreducible representation are given in the Supplementary Material \cite{[{See Supplementary Material at }][{ for character tables with the notation conversion from space group (SG) to point group (PG), the monolayer black P and blue P eigenvectors and a complete classification list of the modes as Raman active, infrared active, acoustic and silent, for all the structures and for $N$ layers.}]SupInfolinkGTPhosphorene}.}} &  [$D^{11}_{2h}$ ($Pbma$, \#57)]~$ 2N(\Gamma^{+}_{1} \oplus \Gamma^{+}_{3} \oplus \Gamma^{-}_{2} \oplus \Gamma^{-}_{4}) \oplus N(\Gamma^{+}_{2} \oplus \Gamma^{+}_{4} \oplus \Gamma^{-}_{1} \oplus \Gamma^{-}_{3})$ \\
\hline
\hline
\multicolumn{3}{c}{Blue P, silicene, germanene and stanene}\\
\hline
   & $AA$ & $AB$  \\
\hline
$N=1$&  &  [$D^{3}_{3d}$ ($P\bar{3}m1$, \#164)]~$\Gamma^{+}_{1} \oplus \Gamma^{+}_{3} \oplus \Gamma^{-}_{2} \oplus \Gamma^{-}_{3}$ \\
$N$ odd ($\neq 1$)&  & [$C^{1}_{3v}$ ($P3m1$, \#156)]~$2N(\Gamma_{1} \oplus \Gamma_{3})$ \\
$N$ even &  \multirow{-4}{*}{[$D^{3}_{3d}$ ($P\bar{3}m1$, \#164)]~$N(\Gamma^{+}_{1} \oplus \Gamma^{+}_{3} \oplus \Gamma^{-}_{2} \oplus \Gamma^{-}_{3})$}  & [$D^{3}_{3d}$ ($P\bar{3}m1$, \#164)]~$N(\Gamma^{+}_{1} \oplus \Gamma^{+}_{3} \oplus \Gamma^{-}_{2} \oplus \Gamma^{-}_{3})$ \\

\end{tabular}
\end{ruledtabular}
\end{sidewaystable}%

The bottom part of Fig.~\ref{correlationgraphene} shows the group-subgroup correlation between the irreducible representations (\emph{IRs}) of the space groups. Some basis functions belonging to each \emph{IR} are also displayed for all the groups. This information is a guide for the analysis of the phase transitions, for example when using infrared ($x,y,z$ basis) or Raman (quadratic basis) spectroscopies. Table\,\ref{tab:Gammavib} displays the irreducible representation decomposition of the vibrational modes ($\Gamma^{vib}$). Excluding the acoustic modes, the remaining irreducible representations for black P monolayer are $2\Gamma_{1}^{+} \oplus \Gamma_{2}^{+} \oplus 2\Gamma_{3}^{+} \oplus \Gamma_{4}^{+} \oplus \Gamma_{1}^{-} \oplus \Gamma_{2}^{-} \oplus \Gamma_{4}^{-}$, and for a blue P monolayer are $\Gamma_{1}^{+} \oplus \Gamma_{3}^{+}$. While in the black P monolayer the one-dimensional representations $\Gamma_{1}^{+}$, $\Gamma_{2}^{+}$, $\Gamma_{3}^{+}$ and $\Gamma_{4}^{+}$ are Raman active, for the blue P monolayer only the $\Gamma_{1}^{+}$ and $\Gamma_{3}^{+}$ [$\Gamma_{3}^{+}$ is doubly degenerate] modes are Raman active. The eigenvectors for black and blue P monolayer are illustrated in Ref. \cite{[{See Supplementary Material at }][{ for character tables with the notation conversion from space group (SG) to point group (PG), the monolayer black P and blue P eigenvectors and a complete classification list of the modes as Raman active, infrared active, acoustic and silent, for all the structures and for $N$ layers.}]SupInfolinkGTPhosphorene}.

    \begin{figure}
        \centering
            \includegraphics[width=0.45\textwidth]{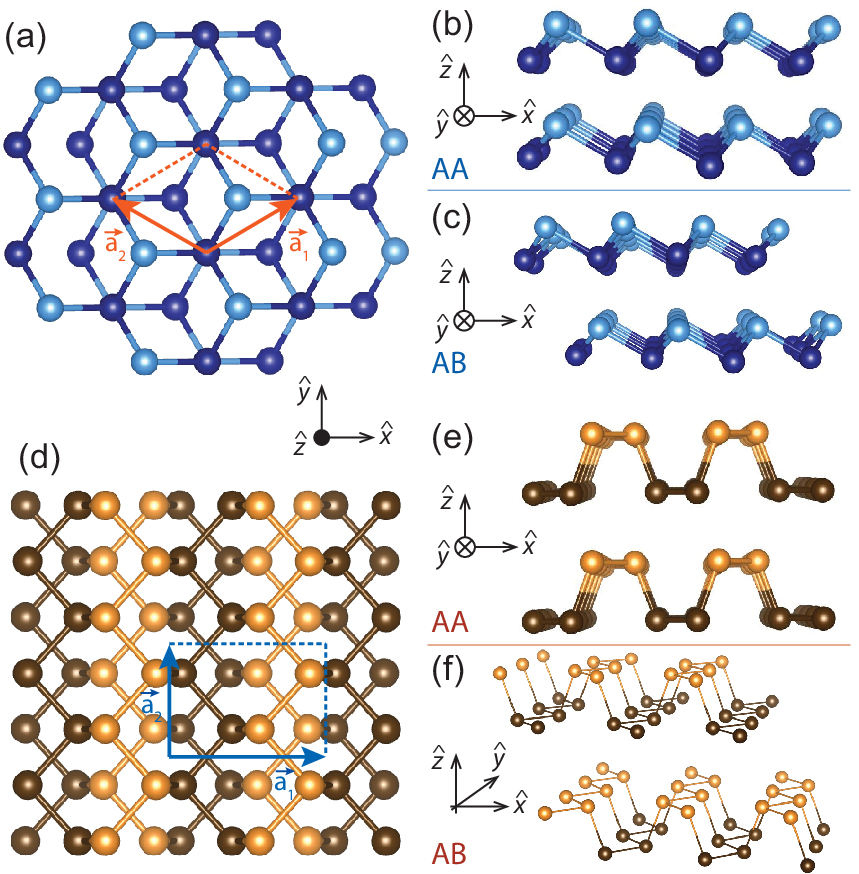}
        \caption{(Color online). Lattice structure of bilayer phosphorene. Color and shading are used to indicate the top and bottom atoms of the non-planar layers. (a) Top view of the $AB$ stacking for blue phosphorus bilayer. (b) and (c): side view of the $AA$ and $AB$ stacking, respectively, for blue phosphorous. (d) Top view of the $AB$ stacking for black phosphorus bilayer. (e) and (f): side view of the $AA$ and $AB$ stacking arrangements, respectively, for black phosphorous bilayer.
        }
        \label{BlackP}
    \end{figure}

The basis functions near each \emph{IR} in the bottom part of Fig.~\ref{correlationgraphene} also guide the polarization dependent analysis in the back and forward Raman scattering configurations. We consider $\hat{z}$ as the light propagation direction, with $\hat{x}, \hat{y}$ as defined in Fig.~\ref{correlationgraphene}. An $xy$ polarization symbol indicates that the polarization of the incident light is in the $x$ direction, and the polarization of the scattered light, in the $y$ direction. For the blue P monolayer, the $\Gamma_{1}^{+}$ mode is detectable under $xx$ and $yy$ polarizations, and the $\Gamma_{3}^{+}$ mode is detectable under $xx$, $yy$ and $xy$ polarizations. In the black P monolayer, the $\Gamma_{1}^{+}$ modes are detectable under $xx$ and $yy$ geometries, while the $\Gamma_{2}^{+}$ mode is detectable in the $xy$ configuration. Therefore, polarization can be used to distinguish black and blue P monolayers from one another. Infrared spectroscopy is also different for the two allotropes: the black P monolayer shows $\Gamma_{2}^{-} \oplus \Gamma_{4}^{-}$ modes that are infrared active, while the blue P does not show any infrared-active mode. In addition, the dependence with polarization can be used to identify the crystallographic orientation of each one of these allotropes.

The \emph{IR} group-subgroup correlations are given by the lines connecting \emph{IRs}. For example, the application of an uniaxial strain to a blue P monolayer, with the $D^{3}_{3d}$ space group, generates a new structure with a $C^{3}_{2h}$ space group symmetry. The optical modes for the $C^{3}_{2h}$ structure are given by $2\Gamma_{1}^{+} \oplus \Gamma_{2}^{+}$, in contrast with the previous $\Gamma_{1}^{+} \oplus \Gamma_{3}^{+}$ of the unstrained structure. The maximum intensity in polarized Raman experiments occurs under the $xy$ configuration for the $\Gamma_{2}^{+}$ mode, and under the $xx$, $yy$, and $xy$ configurations for the $\Gamma_{3}^{+}$ mode. Furthermore, the lowering of symmetry in the strained structure lifts the degeneracy of the $\Gamma_{3}^{+}$ mode, thereby giving rise to two one-dimensional representations. As for the other 2$D$ structures, the change in the peak frequency is expected to provide a measurement of the local strain level of blue phosphorene samples \cite{wang2013raman}.


The space groups and the irreducible representations of the vibrational modes of black and blue phosphorus are repeated in Table \ref{tab:GWVGamma} for the number of layers $N=1$. For a more complete characterization of these systems, it is important to extend the analysis to multiple layers, considering different stacking orders, as displayed in Table \ref{tab:GWVGamma} and in Figure~\ref{BlackP}.

Figs.~\ref{BlackP} (a,c) and (b) show, respectively, the $AB$ and $AA$ stacking arrangements of black P, while Figs. \ref{BlackP} (d,f) and (e) illustrate the corresponding stacks for blue P. The $AA$ stacking for both allotropes occurs when two monolayer units are piled up with each atom of the first monolayer on top of a corresponding atom in the second layer. The top view of the $AA$ stacking arrangement, for both black and blue P, is identical to the monolayer seen in Fig. \ref{correlationgraphene}. In the $AB$ stacking of black P, the second (top) layer is displaced by half of the lattice primitive vector $\vec{a}_{2}$ when comparing with the first (bottom) layer, as shown in Fig. \ref{BlackP} (f). In blue P, an atom of the top layer is placed on top of a non-corresponding atom in the bottom layer, in the other sublattice [see Fig. \ref{BlackP} (a)].

The results presented in Table \ref{tab:GWVGamma} show that different numbers of layers and different stacking arrangements can also result in symmetry variations, and the differences depend on whether $N$ is even or odd. For the $AA$ stacking, an odd and even number $N$ of layers have the same space group ($D^{3}_{3d}$ and $D^{7}_{2h}$ for blue and black P, respectively). The number of modes for $N$ odd and even increases with increasing $N$, following the difference in the number of atoms/unit cell.

In $AB$ black P, the $N$-layered structures can be obtained from exfoliation of the bulk $A17$ phase [$D^{18}_{2h}$ ($Aema$, \#64)], and the space groups for $N$ odd and $N$ even layers are subgroups of the bulk space group. For $AB$ bulk black P, $\Gamma^{vib}=2\Gamma_{1}^{+} \oplus \Gamma_{2}^{+} \oplus 2\Gamma_{3}^{+} \oplus \Gamma_{4}^{+} \oplus \Gamma_{1}^{-} \oplus 2\Gamma_{2}^{-} \oplus \Gamma_{3}^{-} \oplus 2\Gamma_{4}^{-}$ (for $\Gamma^{vib}$ in the standard setting, see Ref. \cite{sugai1985raman}). Only the $\Gamma_{1}^{+}(A_{g})$ ($xx$ and $yy$ polarizations) and $\Gamma_{2}^{+}(B_{1g})$ ($xy$ polarization) modes are Raman active (in the back and forward Raman scattering geometry), and for both $N$ odd and even, these modes correspond to $\Gamma_{1}^{+}$ and $\Gamma_{2}^{+}$, respectively.

On the other hand, for blue P the $AB$ stacking arrangement shows different space groups depending on the number of layers. The $AB$ stacking of blue P is related to the $A7$ phosphorus phase [$D^{5}_{3d}$ ($R\bar{3}m$, \#166) space group, which can be treated as the $ABC$ stacking of $3$ blue P monolayer units]. Once again, it is possible to establish a correlation between the bulk $ABC$ stacking and the bilayer $AB$ stacking (see Ref. \cite{[{See Supplementary Material at }][{ for character tables with the notation conversion from space group (SG) to point group (PG), the monolayer black P and blue P eigenvectors and a complete classification list of the modes as Raman active, infrared active, acoustic and silent, for all the structures and for $N$ layers.}]SupInfolinkGTPhosphorene}). The $\Gamma^{vib}$ for these two systems differs only in the total number of modes due to the change in the number of atoms in the primitive unit cell. Information from Table \ref{tab:GWVGamma} shows a different number of predicted modes and symmetry variations depending on the number of layers for both the $AA$ and $AB$ stacking arrangements of black and blue P with few-layers, and a layer-number dependent comparison analysis can be performed. For $AA$ and $AB$ blue P stacking, as well as for $AA$ black P stacking, to the best of our knowledge, a bulk counterpart has not yet been synthesized.


The presence or absence of inversion is another symmetry-dependent property that can vary with the allotrope, the stacking order, and $N$. In table \ref{tab:GWVGamma} the inversion symmetry is absent only for the $N$ odd ($N>1$) $AB$ stacked blue P $C^{1}_{3v}$ structure. The absence of inversion symmetry in the monolayer version of some Transition Metal Dichalcogenides (TMDs) \cite{Ribeiro-Soares2014GTTMDCs} made it possible to couple spin and valley physics, opening new perspectives for spintronic and valleytronic devices \cite{xiao2012coupled,wang2012electronics}. Furthermore, the absence of inversion symmetry in $N$ odd layers of TMDs have been used in the study of nonlinear optical properties by means of second harmonic generation (SHG) \cite{li2013probing,malard2013observation,zeng2013optical}. In Table \ref{tab:GWVGamma} the structure in which the inversion symmetry is absent is expected to show a significant SHG signal, while the centrosymmetric crystals must show no signal. It is interesting to note that the absence of inversion occurs for $N$ odd layers in the $AB$ blue P (excluding $N=1$). The analysis of the presence vs. absence of the inversion operation for $N$ odd and even layers in the same stacking arrangement based on SHG measurements can, therefore, be used to characterize the crystallographic orientation and number of layers for the $AB$ stacking of blue P.


In summary, we have used group theory to gain insights into the symmetry aspects of black P, blue P, graphene, silicene, germanene and stanene, and their few-layer related systems, in two different stacking arrangements. Our results can be used for a symmetry-based understanding of differences among these systems, and for a fast characterization of in-plane heterostructures that can be built to customize certain desired properties in these new materials. Strained black P and blue P monolayers may exhibit the $C^{1}_{2h}$ subgroup in common. Previous theoretical results suggested a possible conversion trajectory process from black to blue P monolayer with a low activation barrier ($0.47$ eV/atom) \cite{Zhu2014semiconductingblue}, which corresponds to convenient changes in the atomic positions and the stretching of the black P monolayer to obtain blue phosphorene. The group-subgroup relations between black and blue phosphorene corroborate the hypothesis of a mechanical conversion route \cite{Zhu2014semiconductingblue}. The analysis of the irreducible representations of the vibrational modes shows how to distinguish the different systems, and the analysis of inversion symmetry-breaking offers another possibility for identifying the number of layers and their crystallographic orientation, in addition to exploring nonlinear optical phenomena.

We would like to strongly acknowledge Drs. P. Li, I. Appelbaum and A-L. Phaneuf-L'Heureux for a critical reading of the manuscript, and Dr. N. L. Speziali for discussions. The authors acknowledge financial support from FAPEMIG, CNPq grant 551953/2011-0, and NSF grant DMR-$1004147$.

\bibliography{bib}
\addcontentsline{toc}{section}{Referências}

\clearpage

\renewcommand{\thetable}{S \Roman{table}}
\renewcommand{\thefigure}{S\arabic{figure}}

\subsection*{\label{sec:level11} Supplementary Material to ``Group theory for structural analysis and lattice vibrations in phosphorene systems''}

J. Ribeiro-Soares$^{1,}$ $^{2,*}$, R. M. Almeida$^{1}$, L. G. Cançado$^{1}$, M. S. Dresselhaus$^{3}$ and A. Jorio$^{1}$

$^{1}$\textit{Departamento de F\'{\i}sica, Universidade Federal de Minas Gerais, Belo Horizonte, MG, 30123-970, Brazil}

$^{2}$\textit{Departamento de F\'{\i}sica, Universidade Federal de Lavras, Lavras, MG, 37200-000, Brazil}

$^{3}$\textit{Department of Electrical Engineering and Computer Science, Massachusetts Institute of Technology (MIT), Cambridge, MA 02139, USA and Department of Physics, Massachusetts Institute of Technology (MIT), Cambridge, MA 02139, USA}\\

*Author to whom correspondence should be addressed: jenainassoares2@gmail.com
\clearpage

\setcounter{table}{0}
\setcounter{figure}{0}

\section*{\label{sec:level1}Contents}
 \begin{enumerate}

   \item Monolayer black P and blue P eigenvectors

   \item Character tables with space group (SG) to point group (PG) notation conversion with convenient basis functions and mode activity classification list.
    \begin{enumerate}

      \item Character table for $D_{6h}^{1}$ and $D_{2h}^{19}$ space groups
            \begin{enumerate}
             \item $D_{6h}^{1}$;
             \item $D_{2h}^{19}$.
            \end{enumerate}

      \item Blue phosphorus monolayer under strain ($C_{2h}^{1}$ and $C_{2h}^{3}$ space groups)

      \item Black phosphorus for $N$ odd and $N$ even number of layers ($D_{2h}^{7}$ and $D_{2h}^{11}$ space groups)
           \begin{enumerate}
            \item $AA$ stacking for $N$ even and $N$ odd, and $AB$ stacking for $N$ odd;
            \item $AB$ stacking for $N$ even.
           \end{enumerate}
      \item Blue phosphorus, silicene germanene and stanene for $N$ odd and $N$ even number of layers ($D_{3d}^{3}$ and $C_{3v}^{1}$ space groups)
            \begin{enumerate}
            \item $AA$ stacking for $N$ even and $N$ odd, and $AB$ stacking for monolayer and $N$ even;
            \item $AB$ stacking for $N$ odd.
           \end{enumerate}

      \item Bulk counterparts
            \begin{enumerate}
            \item $A17$ phase;
            \item $A7$ phase (ABC stacking and relation with $AB$ blue phosphorus bilayer).
            \end{enumerate}
    \end{enumerate}

   \end{enumerate}

\clearpage

\begin{enumerate}

\item Monolayer black P and blue P eigenvectors.

   \begin{figure}[h!]
        \centering
            \includegraphics[width=1\textwidth]{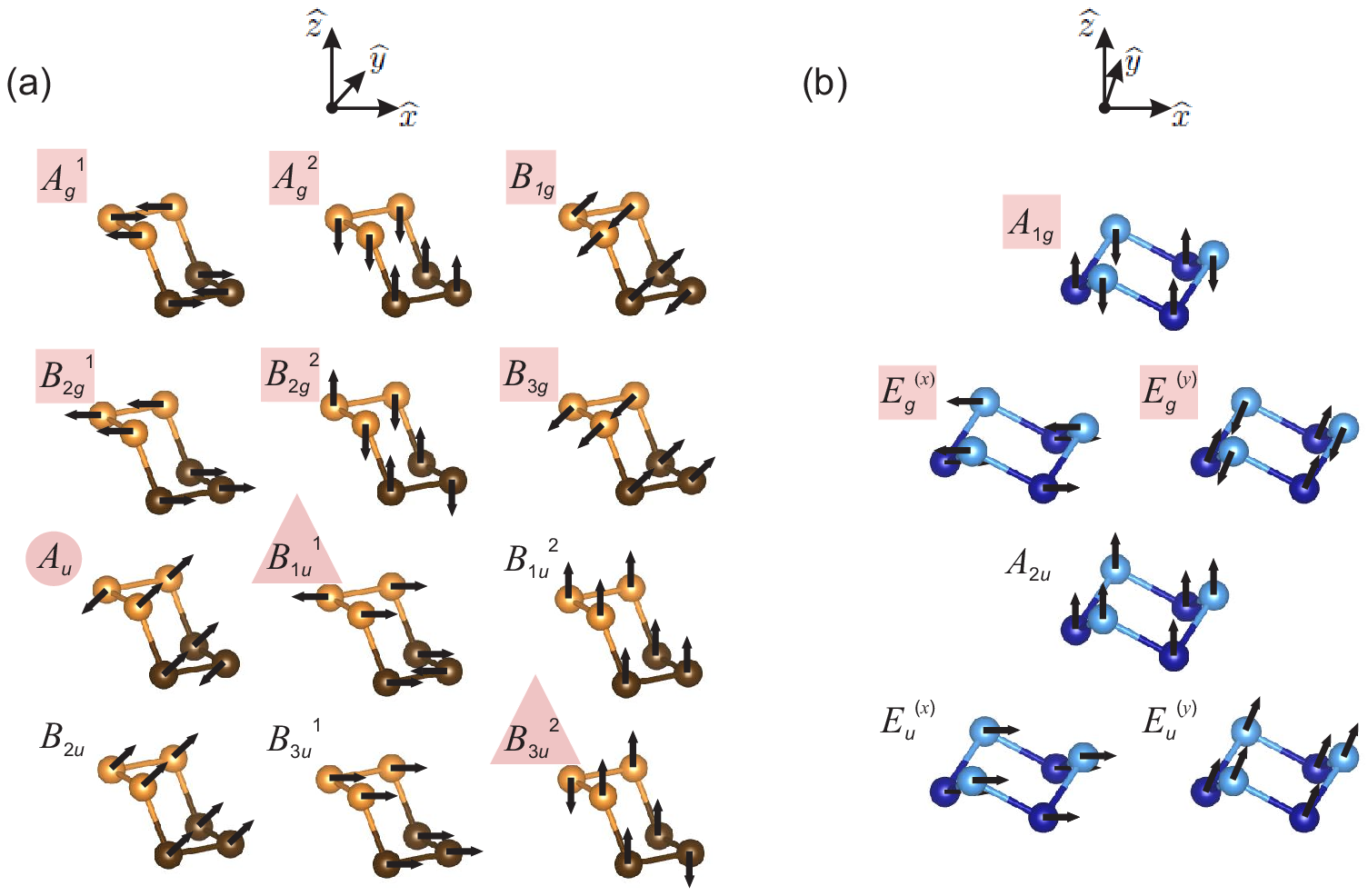}
        \caption{Non-normalized eigenvector representations for black (a) and blue (b) P monolayer vibrational modes, in its respective axis choices. Pink rectangles indicate Raman active modes, pink triangles represent infrared active modes and pink circles indicate silent modes. The remaining modes are acoustic.}
        \label{eigenvecblackblue}
    \end{figure}
\FloatBarrier

   \item Character tables with space group (SG) to point group (PG) notation conversion with convenient basis functions and mode activity classification list.
    \begin{enumerate}

    \item Character table for $D_{6h}^{1}$ and $D_{2h}^{19}$ space groups
            \begin{enumerate}
             \item $D_{6h}^{1}$;

   \begin{table*}[!ht]
 \caption[]{Character table for the $\Gamma$ point [$D_{6h}^{1}$ ($P6/mmm$, \#191)].}\vspace{0.3cm}\footnotesize\label{chargamma2}
 \begin{ruledtabular}
\begin{tabular}{ccccccccccccccc}
   &&  &  &  &  & $C_{2}^{'A}$ & $C_{2}^{''A}$ &  &  &  &  & $\sigma_{d}^{A}$ & $\sigma_{v}^{A}$ &    \\
   &&  & $C_{3}^{+}$ &  & $C_{6}^{-}$ & $C_{2}^{'B}$ & $C_{2}^{''B}$ &  & $S_{6}^{+}$ &  & $S_{3}^{-}$ & $\sigma_{d}^{B}$ & $\sigma_{v}^{B}$ &    \\
   SG&PG& $E$ & $C_{3}^{-}$ & $C_{2}$ & $C_{6}^{+}$ & $C_{2}^{'C}$ & $C_{2}^{''C}$ & i & $S_{6}^{-}$ & $\sigma_{h}$ & $S_{3}^{+}$ & $\sigma_{d}^{C}$ & $\sigma_{v}^{C}$ &  Basis  \\
\hline
$\Gamma_{1}^{+}$ & $A_{1g}$ &  1 & 1 & 1 & 1 & 1 & 1 & 1 & 1 & 1 & 1 & 1 & 1 &  $x^{2}+y^{2}\;,\;z^{2}$  \\
$\Gamma_{2}^{+}$ & $A_{2g}$ &  1 & 1 & 1 & 1 & -1 & -1 & 1 & 1 & 1 & 1 & -1 & -1 &   \\
$\Gamma_{3}^{+}$ & $B_{1g}$ & 1 & 1 & -1 & -1 & -1 & 1 & 1 & 1 & -1 & -1 & -1 & 1 &   \\
$\Gamma_{4}^{+}$ & $B_{2g}$ & 1 & 1 & -1 & -1 & 1 & -1 & 1 & 1 & -1 & -1 & 1 & -1 &   \\
$\Gamma_{5}^{+}$ & $E_{1g}$ & 2 & -1 & -2 & 1 & 0 & 0 & 2 & -1 & -2 & 1 & 0 & 0 &  $(xz,yz)$ \\
$\Gamma_{6}^{+}$ & $E_{2g}$ & 2 & -1 & 2 & -1 & 0 & 0 & 2 & -1 & 2 & -1 & 0 & 0 &  $(x^{2}-y^{2},xy)$ \\
$\Gamma_{1}^{-}$ & $A_{1u}$ & 1 & 1 & 1 & 1 & 1 & 1 & -1 & -1 & -1 & -1 & -1 & -1 &   \\
$\Gamma_{2}^{-}$ & $A_{2u}$ & 1 & 1 & 1 & 1 & -1 & -1 & -1 & -1 & -1 & -1 & 1 & 1 &  $z$ \\
$\Gamma_{3}^{-}$ & $B_{1u}$ & 1 & 1 & -1 & -1 & -1 & 1 & -1 & -1 & 1 & 1 & 1 & -1 &   \\
$\Gamma_{4}^{-}$ & $B_{2u}$ & 1 & 1 & -1 & -1 & 1 & -1 & -1 & -1 & 1 & 1 & -1 & 1 &   \\
$\Gamma_{5}^{-}$ & $E_{1u}$ & 2 & -1 & -2 & 1 & 0 & 0 & -2 & 1 & 2 & -1 & 0 & 0 &  $(x,y)$ \\
$\Gamma_{6}^{-}$ & $E_{2u}$ &2 & -1 & 2 & -1 & 0 & 0 & -2 & 1 & -2 & 1 & 0 & 0 &   \\
\end{tabular}
\end{ruledtabular}
\end{table*}

\begin{table}[h!]
\caption{\label{tab:RamanIRACSilbulk} Normal vibrational mode irreducible representations ($\Gamma^{vib}$) for graphene. Irreducible representations for the Raman active, infrared active, acoustic and silent mode are identified.}

\begin{tabular}{cc}
\hline
\hline
 \multicolumn{2}{c}{$D_{6h}^{1}$ ($P6/mmm$, \#191)} \\
\hline
$\Gamma^{vib}$ & $\Gamma_{4}^+ \oplus \Gamma_{6}^+ \oplus \Gamma_{2}^- \oplus \Gamma_{5}^-$ \\
\hline
Raman & $\Gamma_{6}^+$   \\
Infrared & -  \\
Acoustic & $\Gamma_{2}^- \oplus \Gamma_{5}^-$  \\
Silent & $\Gamma_{4}^+$ \\
\hline
\hline
\end{tabular}
\end{table}

\FloatBarrier

            \item $D_{2h}^{19}$.

\begin{table*}[h!]
\caption{\label{tab:D2hstrainedgraphene_black} Character table for the $\Gamma$ point [$D^{19}_{2h}$ ($Cmmm$, \#65)]. The symmetry operations associated to the centring translation [centring vector: ($\frac{1}{2},\frac{1}{2},0$) ] are not shown.}
\vspace{0.3cm}\footnotesize
\begin{ruledtabular}
\begin{tabular}{ccccccccccc}

  SG & PG & $\{E|0\}$ & $\{C_{2\hat{z}}|0\}$ & $\{C_{2\hat{y}}|0\}$ & $\{C_{2\hat{x}}|0\}$ & $\{i|0\}$ & $\{\sigma_{xy}|0\}$ & $\{\sigma_{xz}|0\}$ & $\{\sigma_{yz}|0\}$ & Basis \\
\hline
 $\Gamma^{+}_{1}$ & $A_{g}$ & $1$ & $1$ & $1$ & $1$ & $1$ & $1$ & $1$ & $1$ & $x^{2}, y^{2}, z^{2}$ \\
 $\Gamma^{+}_{2}$ & $B_{1g}$ & $1$ & $1$ & $-1$ & $-1$ & $1$ & $1$ & $-1$ & $-1$ & $ xy$ \\
 $\Gamma^{+}_{3}$ & $B_{2g}$ & $1$ & $-1$ & $1$ & $-1$ & $1$ & $-1$ & $1$ & $-1$ & $ xz$ \\
 $\Gamma^{+}_{4}$ & $B_{3g}$ & $1$ & $-1$ & $-1$ & $1$ & $1$ & $-1$ & $-1$ & $1$ & $ yz$ \\
 $\Gamma^{-}_{1}$ & $A_{u}$ & $1$ & $1$ & $1$ & $1$ & $-1$ & $-1$ & $-1$ & $-1$ & \\
 $\Gamma^{-}_{2}$ & $B_{1u}$ & $1$ & $1$ & $-1$ & $-1$ & $-1$ & $-1$ & $1$ & $1$ & $z$ \\
 $\Gamma^{-}_{3}$ & $B_{2u}$ & $1$ & $-1$ & $1$ & $-1$ & $-1$ & $1$ & $-1$ & $1$ & $y$ \\
 $\Gamma^{-}_{4}$ & $B_{3u}$ & $1$ & $-1$ & $-1$ & $1$ & $-1$ & $1$ & $1$ & $-1$ & $x$ \\
\end{tabular}
\end{ruledtabular}
    \end{table*}

\begin{table}[h!]
\caption{\label{tab:RamanIRACSil_strainedgraphene} Normal vibrational mode irreducible representations ($\Gamma^{vib}$) for strained graphene. Irreducible representations for the Raman active, infrared active, acoustic and silent mode are identified.}

\begin{tabular}{cc}
\hline
\hline
 \multicolumn{2}{c}{$D^{19}_{2h}$ ($Cmmm$, \#65)} \\
\hline
$\Gamma^{vib}$ & $ \Gamma^{+}_{1} \oplus \Gamma^{+}_{2} \oplus \Gamma^{+}_{3} \oplus \Gamma^{-}_{2} \oplus \Gamma^{-}_{3} \oplus \Gamma^{-}_{4} $ \\
\hline
Raman & $\Gamma^{+}_{1} \oplus \Gamma^{+}_{2} \oplus \Gamma^{+}_{3}$  \\
Infrared & -  \\
Acoustic & $\Gamma^{-}_{2} \oplus \Gamma^{-}_{3} \oplus \Gamma^{-}_{4}$ \\
Silent & - \\
\hline
\hline
\end{tabular}
\end{table}

            \end{enumerate}

\FloatBarrier

\item Blue phosphorus monolayer under strain ($C_{2h}^{1}$ and $C_{2h}^{3}$ space groups)

 \begin{table*}[h!]
\caption{\label{tab:C2hstrain_blueP} Character table for the $\Gamma$ point [$C^{1}_{2h}$ ($P2/m$, \#10)].}
\begin{ruledtabular}
\begin{tabular}{ccccccc}
 SG & PG & $\{E|0\}$ & $\{C_{2\hat{y}}|0\}$ & $\{\sigma_{xz}|0\}$ & $\{i|0\}$ & Basis \\
\hline
 $\Gamma^{+}_{1}$ & $A_{g}$ & $1$ & $1$ & $1$ & $1$ & $x^{2}, y^{2}, z^{2}, xz$\\
 $\Gamma^{-}_{1}$ & $A_{u}$ & $1$ & $1$ & $-1$ & $-1$ & $y$ \\
 $\Gamma^{+}_{2}$ & $B_{g}$ & $1$ & $-1$ & $-1$ & $1$  & $xy, yz$\\
 $\Gamma^{-}_{2}$ & $B_{u}$ & $1$ & $-1$ & $1$ & $-1$ & $x, z$\\
\end{tabular}
\end{ruledtabular}
    \end{table*}

The character table to the $C^{1}_{2h}$ group is used to the $C^{3}_{2h}$ ($C2/m$, \#12) group, in which additional symmetry operations associated to a centring translation [centring vector: ($\frac{1}{2},\frac{1}{2},0$)] occurs, but the point group is the same for both $C^{1}_{2h}$ and $C^{3}_{2h}$.

\begin{table}[h!]
\caption{\label{tab:RamanIRACSilbulk} Normal vibrational mode irreducible representations ($\Gamma^{vib}$) for strained (or stressed) Blue P monolayer at the $\Gamma$ point ($C^{3}_{2h}$). Irreducible representations for the Raman active, infrared active, acoustic and silent mode are identified.}

\begin{tabular}{cc}
\hline
\hline
 \multicolumn{2}{c}{$C^{3}_{2h}$ ($C2/m$, \#12)} \\
\hline
$\Gamma^{vib}$ & $2\Gamma_{1}^+ \oplus 2\Gamma_{2}^- \oplus \Gamma_{1}^- \oplus \Gamma_{2}^+$ \\
\hline
Raman & $2\Gamma_{1}^+ \oplus \Gamma_{2}^+$   \\
Infrared & -  \\
Acoustic & $\Gamma_{1}^- \oplus 2\Gamma_{2}^-$  \\
Silent & - \\
\hline
\hline
\end{tabular}
\end{table}

\begin{table}[h!]
\caption{\label{tab:RamanIRACSilbulk} Normal vibrational mode irreducible representations ($\Gamma^{vib}$) for strained (or stressed) and distorted Blue P monolayer at the $\Gamma$ point ($C^{1}_{2h}$). Irreducible representations for the Raman active, infrared active, acoustic and silent mode are identified.}

\begin{tabular}{cc}
\hline
\hline
 \multicolumn{2}{c}{$C^{1}_{2h}$ ($P2/m$, \#10)} \\
\hline
$\Gamma^{vib}$ & $4\Gamma_{1}^+ \oplus 4\Gamma_{2}^- \oplus 2\Gamma_{1}^- \oplus 2\Gamma_{2}^+$ \\
\hline
Raman & $4\Gamma_{1}^+ \oplus 2\Gamma_{2}^+$   \\
Infrared & $2\Gamma_{2}^- \oplus 1\Gamma_{1}^-$  \\
Acoustic & $2\Gamma_{2}^- \oplus 1\Gamma_{1}^-$  \\
Silent & - \\
\hline
\hline
\end{tabular}
\end{table}

\FloatBarrier

      \item Black phosphorus for $N$ odd and $N$ even number of layers ($D_{2h}^{7}$ and $D_{2h}^{11}$ space groups)
           \begin{enumerate}
            \item $AA$ stacking for $N$ even and $N$ odd, and $AB$ stacking for $N$ odd;

 \begin{table*}[h!]
\caption{\label{tab:D2hmonolayer_black} Character table for the $\Gamma$ point [$D^{7}_{2h}$ ($Pbmn$, \#53)].}
\vspace{0.3cm}\footnotesize
\begin{ruledtabular}
\begin{tabular}{ccccccccccc}

  SG & PG & $\{E|0\}$ & $\{C_{2\hat{z}(x=y=1/4)}|0\}$ & $\{C_{2\hat{y}}|0\}$ & $\{C_{2\hat{x}(y=\frac{1}{4})}|\tau_{x}\}$\footnote{$\tau_{x}$ is the translation of half of the $a_{1}$ lattice parameter along the $\hat{x}$ direction [$\tau_{x}=(\frac{1}{2})a_{1}\hat{x}$].\label{fn:repeattaux}} & $\{i|0\}$ & $\{\sigma_{xy}|\tau_{n}\}$\footnote{$\tau_{n}$ is the translation of half of the $a_{1}$ lattice parameter along the $\hat{x}$ direction and the translation of half of the $a_{2}$ lattice parameter along the $\hat{y}$ direction [$\tau_{n}=(\frac{1}{2})a_{1}\hat{x}+(\frac{1}{2})a_{2}\hat{y}$].\label{fn:repeattaun}} & $\{\sigma_{xz}|0\}$ & $\{\sigma_{yz(x=\frac{1}{4})}|\tau_{y}\}$\footnote{$\tau_{y}$ is the translation of half of the $a_{2}$ lattice parameter along the $\hat{y}$ direction [$\tau_{y}=(\frac{1}{2})a_{2}\hat{y}$].\label{fn:repeattauy}} & Basis \\
\hline
 $\Gamma^{+}_{1}$ & $A_{g}$ & $1$ & $1$ & $1$ & $1$ & $1$ & $1$ & $1$ & $1$ & $x^{2}, y^{2}, z^{2}$ \\
 $\Gamma^{+}_{2}$ & $B_{1g}$ & $1$ & $1$ & $-1$ & $-1$ & $1$ & $1$ & $-1$ & $-1$ & $ xy$ \\
 $\Gamma^{+}_{3}$ & $B_{2g}$ & $1$ & $-1$ & $1$ & $-1$ & $1$ & $-1$ & $1$ & $-1$ & $ xz$ \\
 $\Gamma^{+}_{4}$ & $B_{3g}$ & $1$ & $-1$ & $-1$ & $1$ & $1$ & $-1$ & $-1$ & $1$ & $ yz$ \\
 $\Gamma^{-}_{1}$ & $A_{u}$ & $1$ & $1$ & $1$ & $1$ & $-1$ & $-1$ & $-1$ & $-1$ & \\
 $\Gamma^{-}_{2}$ & $B_{1u}$ & $1$ & $1$ & $-1$ & $-1$ & $-1$ & $-1$ & $1$ & $1$ & $z$ \\
 $\Gamma^{-}_{3}$ & $B_{2u}$ & $1$ & $-1$ & $1$ & $-1$ & $-1$ & $1$ & $-1$ & $1$ & $y$ \\
 $\Gamma^{-}_{4}$ & $B_{3u}$ & $1$ & $-1$ & $-1$ & $1$ & $-1$ & $1$ & $1$ & $-1$ & $x$ \\
\end{tabular}
\end{ruledtabular}
    \end{table*}

\begin{table}[h!]
\caption{\label{tab:RamanIRACSil_black_AA_ABNodd} Normal vibrational mode irreducible representations ($\Gamma^{vib}$) for black phosphorus $AA$ stacking ($N$ even and $N$ odd), and $AB$ stacking ($N$ odd) at the $\Gamma$ point. Irreducible representations for the Raman active, infrared active, acoustic and silent mode are identified.}

\begin{tabular}{cc}
\hline
\hline
 \multicolumn{2}{c}{$D^{7}_{2h}$ ($Pbmn$, \#53)} \\
\hline
$\Gamma^{vib}$ & $ 2N(\Gamma^{+}_{1} \oplus \Gamma^{+}_{3} \oplus \Gamma^{-}_{2} \oplus \Gamma^{-}_{4}) \oplus N(\Gamma^{+}_{2} \oplus \Gamma^{+}_{4} \oplus \Gamma^{-}_{1} \oplus \Gamma^{-}_{3})$ \\
\hline
Raman & $2N(\Gamma^{+}_{1} \oplus \Gamma^{+}_{3}) \oplus N(\Gamma^{+}_{2} \oplus \Gamma^{+}_{4})$  \\
Infrared & $(2N-1)(\Gamma_{2}^- \oplus \Gamma_{4}^-) \oplus (N-1)\Gamma_{3}^- $  \\
Acoustic & $\Gamma_{2}^- \oplus \Gamma_{3}^- \oplus \Gamma_{4}^-$ \\
Silent & $N\Gamma_{1}^-$ \\
\hline
\hline
\end{tabular}
\end{table}
\FloatBarrier

            \item $AB$ stacking for $N$ even;

 \begin{table*}[h!]
 \centering
\caption{\label{tab:D2hbilayer_black} Character table for the $\Gamma$ point [$D^{11}_{2h}$ ($Pbma$, \#57)].}
\vspace{0.3cm}\scriptsize
\begin{ruledtabular}
\begin{tabular}{ccccccccccc}
  SG & PG & $\{E|0\}$ & $\{C_{2\hat{z}(x=\frac{1}{4})}|0\}$ & $\{C_{2\hat{y}}|\tau_{y}\}$\footnote{$\tau_{y}$ is the translation of half of the $a_{2}$ lattice parameter along the $\hat{y}$ direction [$\tau_{y}=(\frac{1}{2})a_{2}\hat{y}$].\label{fn:repeattauy}} & $\{C_{2\hat{x}(y=\frac{1}{4})}|\tau_{x}\}$\footnote{$\tau_{x}$ is the translation of half of the $a_{1}$ lattice parameter along the $\hat{x}$ direction [$\tau_{x}=(\frac{1}{2})a_{1}\hat{x}$].\label{fn:repeattaux}} & $\{i|0\}$ & $\{\sigma_{xy}|\tau_{x}\}$\footref{fn:repeattaux} & $\{\sigma_{xz(y=\frac{1}{4})}|0\}$ & $\{\sigma_{yz(x=\frac{1}{4})}|\tau_{y}\}$\footref{fn:repeattauy} & Basis \\
\hline
 $\Gamma^{+}_{1}$ & $A_{g}$ & $1$ & $1$ & $1$ & $1$ & $1$ & $1$ & $1$ & $1$ & $x^{2}, y^{2}, z^{2}$ \\
 $\Gamma^{+}_{2}$ & $B_{1g}$ & $1$ & $1$ & $-1$ & $-1$ & $1$ & $1$ & $-1$ & $-1$ & $ xy$ \\
 $\Gamma^{+}_{3}$ & $B_{2g}$ & $1$ & $-1$ & $1$ & $-1$ & $1$ & $-1$ & $1$ & $-1$ & $ xz$ \\
 $\Gamma^{+}_{4}$ & $B_{3g}$ & $1$ & $-1$ & $-1$ & $1$ & $1$ & $-1$ & $-1$ & $1$ & $ yz$ \\
 $\Gamma^{-}_{1}$ & $A_{u}$ & $1$ & $1$ & $1$ & $1$ & $-1$ & $-1$ & $-1$ & $-1$ & \\
 $\Gamma^{-}_{2}$ & $B_{1u}$ & $1$ & $1$ & $-1$ & $-1$ & $-1$ & $-1$ & $1$ & $1$ & $z$ \\
 $\Gamma^{-}_{3}$ & $B_{2u}$ & $1$ & $-1$ & $1$ & $-1$ & $-1$ & $1$ & $-1$ & $1$ & $y$ \\
 $\Gamma^{-}_{4}$ & $B_{3u}$ & $1$ & $-1$ & $-1$ & $1$ & $-1$ & $1$ & $1$ & $-1$ & $x$ \\
\end{tabular}
%
\end{ruledtabular}
    \end{table*}

\begin{table}[h!]
\begin{center}
\caption{\label{tab:RamanIRACSil_black_AA_ABNeven} Normal vibrational mode irreducible representations ($\Gamma^{vib}$) for black phosphorus $AB$ stacking ($N$ even) at the $\Gamma$ point. Irreducible representations for the Raman active, infrared active, acoustic and silent mode are identified.}

\begin{tabular}{cc}
\hline
\hline
 \multicolumn{2}{c}{$D^{11}_{2h}$ ($Pbma$, \#57)} \\
\hline
$\Gamma^{vib}$ & $ 2N(\Gamma^{+}_{1} \oplus \Gamma^{+}_{3} \oplus \Gamma^{-}_{2} \oplus \Gamma^{-}_{4}) \oplus N(\Gamma^{+}_{2} \oplus \Gamma^{+}_{4} \oplus \Gamma^{-}_{1} \oplus \Gamma^{-}_{3})$ \\
\hline
Raman & $2N(\Gamma^{+}_{1} \oplus \Gamma^{+}_{3}) \oplus N(\Gamma^{+}_{2} \oplus \Gamma^{+}_{4})$  \\
Infrared & $(2N-1)(\Gamma_{2}^- \oplus \Gamma_{4}^-) \oplus (N-1)\Gamma_{3}^- $  \\
Acoustic & $\Gamma_{2}^- \oplus \Gamma_{3}^- \oplus \Gamma_{4}^-$ \\
Silent & $N\Gamma_{1}^-$ \\
\hline
\hline
\end{tabular}
\end{center}
\end{table}
\FloatBarrier

           \end{enumerate}

\FloatBarrier
      \item Blue phosphorus, silicene, germanene and stanene for $N$ odd and $N$ even number of layers ($D_{3d}^{3}$ and $C_{3v}^{1}$ space groups)
            \begin{enumerate}
            \item $AA$ stacking for $N$ even and $N$ odd, and $AB$ stacking for monolayer and $N$ even;
\begin{table*}[h!]
\caption{\label{tab:D3deven} Character table for the $\Gamma$ point [$D^{3}_{3d}$ ($P\bar{3}m1$, \#164)].}
\begin{ruledtabular}
\begin{tabular}{ccccccccc}
  &  &  &  & $C'^A_{2}$ &  &  & $\sigma^A_{d}$ &  \\
  &  &  & $C^+_{3}$ & $C'^B_{2}$ &  & $S^+_{6}$ & $\sigma^B_{d}$ &  \\
 SG & PG & $E$ & $C^-_{3}$ & $C'^C_{2}$ & $i$ & $S^-_{6}$ & $\sigma^C_{d}$ & Basis \\
\hline
$\Gamma^{+}_{1}$ & $A_{1g}$ & $1$ & $1$ & $1$ & $1$ & $1$ & $1$ & $x^{2}+y^{2}, z^{2}$ \\
$\Gamma^{+}_{2}$ & $A_{2g}$ & $1$ & $1$ & $-1$ & $1$ & $1$ & $-1$ &  \\
$\Gamma^{+}_{3}$ & $E_{g}$ & $2$ & $-1$ & $0$ & $2$ & $-1$ & $0$ & $(xz, yz), (x^{2}-y^{2}, xy)$ \\
$\Gamma^{-}_{1}$ & $A_{1u}$ & $1$ & $1$ & $1$ & $-1$ & $-1$ & $-1$ &  \\
$\Gamma^{-}_{2}$ & $A_{2u}$ & $1$ & $1$ & $-1$ & $-1$ & $-1$ & $1$ & $z$ \\
$\Gamma^{-}_{3}$ & $E_{u}$ & $2$ & $-1$ & $0$ & $-2$ & $1$ & $0$ & $(x, y)$ \\
\end{tabular}
\end{ruledtabular}
    \end{table*}

\begin{table}[h!]
\caption{\label{tab:RamanIRACSilbulk} Normal vibrational mode irreducible representations ($\Gamma^{vib}$) for blue phosphorus, silicene, germanene and stanene $AA$ stacking ($N$ even and $N$ odd), and $AB$ stacking (monolayer and $N$ even) at the $\Gamma$ point. Irreducible representations for the Raman active, infrared active, acoustic and silent mode are identified.}

\begin{tabular}{cc}
\hline
\hline
 \multicolumn{2}{c}{$D^{3}_{3d}$ ($P\bar{3}m1$, \#164)} \\
\hline
$\Gamma^{vib}$ & $N(\Gamma_{1}^+ \oplus \Gamma_{3}^+ \oplus \Gamma_{2}^- \oplus \Gamma_{3}^-)$ \\
\hline
Raman & $N(\Gamma_{1}^+ \oplus \Gamma_{3}^+)$   \\
Infrared & $(N-1)(\Gamma_{2}^- \oplus \Gamma_{3}^-)$  \\
Acoustic & $\Gamma_{2}^- \oplus \Gamma_{3}^-$  \\
Silent & - \\
\hline
\hline
\end{tabular}
\end{table}
\FloatBarrier

            \item $AB$ stacking for $N$ odd;

\begin{table*}[h!]
\caption{\label{tab:D3even} Character table for the $\Gamma$ point [$C^{1}_{3v}$ ($P3m1$, \#156)].}
\begin{ruledtabular}
\begin{tabular}{cccccc}
  &  &  &  & $\sigma^A_{d}$ &  \\
  &  &  & $C^+_{3}$ & $\sigma^B_{d}$ &  \\
 SG & PG & $E$ & $C^-_{3}$ & $\sigma^C_{d}$ & Basis \\
\hline
 $\Gamma_{1}$ & $A_{1}$ & $1$ & $1$ & $1$ & $z, x^{2}+y^{2}, z^{2}$ \\
 $\Gamma_{2}$ & $A_{2}$ & $1$ & $1$ & $-1$ &  \\
 \multirow{2}{*}{$\Gamma_{3}$} & \multirow{2}{*}{$E$} & \multirow{2}{*}{$2$} & \multirow{2}{*}{$-1$} & \multirow{2}{*}{$0$} & \multirow{2}{*}{$(xz, yz), (x, y)$} \\
  &  &  &  & & $(x^{2}-y^{2}, xy)$ \\
\end{tabular}
\end{ruledtabular}
\end{table*}

\begin{table}[h!]
\caption{\label{tab:RamanIRACSilbulk} Normal vibrational mode irreducible representations ($\Gamma^{vib}$) for blue phosphorus, silicene, germanene and stanene $AB$ stacking ($N$ odd) at the $\Gamma$ point. Irreducible representations for the Raman active, infrared active, acoustic and silent mode are identified.}

\begin{tabular}{cc}
\hline
\hline
 \multicolumn{2}{c}{$C^{1}_{3v}$ ($P3m1$, \#156)} \\
\hline
$\Gamma^{vib}$ & $~~~2N(\Gamma_{1} \oplus \Gamma_{3})$ \\
\hline
Raman & $~~~(2N-1)(\Gamma_{1} \oplus \Gamma_{3})$  \\
Infrared & $~~~(2N-1)(\Gamma_{1} \oplus \Gamma_{3})$  \\
Acoustic & $~~~\Gamma_{1} \oplus \Gamma_{3}$ \\
Silent & - \\
\hline
\hline
\end{tabular}
\end{table}
\FloatBarrier

           \end{enumerate}

      \item Bulk counterparts
            \begin{enumerate}
            \item $A17$ phase [black phosphorus, $D^{18}_{2h}$ ($Aema$, \#64)];
\begin{table*}[h!]
\caption{\label{tab:D2hbulk} Character table for the $\Gamma$ point [$D^{18}_{2h}$ ($Aema$, \#64)]. The symmetry operations associated to the centring translation [centring vector: ($0,\frac{1}{2},\frac{1}{2}$) ] are not shown.}
\vspace{0.3cm}\scriptsize
\begin{ruledtabular}
\begin{tabular}{ccccccccccc}

  SG & PG & $\{E|0\}$ & $\{C_{2\hat{z}(x=\frac{1}{4})}|\tau_{z}\}$\footnote{$\tau_{z}$ is the translation of half of the $c$ lattice parameter along the $\hat{z}$ direction [$\tau_{z}=(\frac{1}{2})c\hat{z}$].\label{fn:repeattauz}} & $\{C_{2\hat{y}}|0\}$ & $\{C_{2\hat{x}(z=\frac{1}{4})}|\tau_{x}\}$\footnote{$\tau_{x}$ is the translation of half of the $a_{1}$ lattice parameter along the $\hat{x}$ direction [$\tau_{x}=(\frac{1}{2})a_{1}\hat{x}$].\label{fn:repeattaux}} & $\{i|0\}$ & $\{\sigma_{xy(z=1/4)}|\tau_{x}\}$\footref{fn:repeattaux} & $\{\sigma_{xz}|0\}$ & $\{\sigma_{yz(x=\frac{1}{4})}|\tau_{z}\}$\footref{fn:repeattauz} & Basis \\
\hline
 $\Gamma^{+}_{1}$ & $A_{g}$ & $1$ & $1$ & $1$ & $1$ & $1$ & $1$ & $1$ & $1$ & $x^{2}, y^{2}, z^{2}$ \\
 $\Gamma^{+}_{2}$ & $B_{1g}$ & $1$ & $1$ & $-1$ & $-1$ & $1$ & $1$ & $-1$ & $-1$ & $ xy$ \\
 $\Gamma^{+}_{3}$ & $B_{2g}$ & $1$ & $-1$ & $1$ & $-1$ & $1$ & $-1$ & $1$ & $-1$ & $ xz$ \\
 $\Gamma^{+}_{4}$ & $B_{3g}$ & $1$ & $-1$ & $-1$ & $1$ & $1$ & $-1$ & $-1$ & $1$ & $ yz$ \\
 $\Gamma^{-}_{1}$ & $A_{u}$ & $1$ & $1$ & $1$ & $1$ & $-1$ & $-1$ & $-1$ & $-1$ & \\
 $\Gamma^{-}_{2}$ & $B_{1u}$ & $1$ & $1$ & $-1$ & $-1$ & $-1$ & $-1$ & $1$ & $1$ & $z$ \\
 $\Gamma^{-}_{3}$ & $B_{2u}$ & $1$ & $-1$ & $1$ & $-1$ & $-1$ & $1$ & $-1$ & $1$ & $y$ \\
 $\Gamma^{-}_{4}$ & $B_{3u}$ & $1$ & $-1$ & $-1$ & $1$ & $-1$ & $1$ & $1$ & $-1$ & $x$ \\
\end{tabular}
\end{ruledtabular}
    \end{table*}

\begin{table}[h!]
\caption{\label{tab:RamanIRACSilbulk} Normal vibrational mode irreducible representations ($\Gamma^{vib}$) for $A17$ phosphorus phase. Irreducible representations for the Raman active, infrared active, acoustic and silent mode are identified.}

\begin{tabular}{cc}
\hline
\hline
 \multicolumn{2}{c}{$D^{18}_{2h}$ ($Aema$, \#64)} \\
\hline
$\Gamma^{vib}$ & $2\Gamma_{1}^+ \oplus \Gamma_{2}^+ \oplus 2\Gamma_{3}^+ \oplus \Gamma_{4}^+ \oplus \Gamma_{1}^- \oplus 2\Gamma_{2}^- \oplus \Gamma_{3}^- \oplus 2\Gamma_{4}^-$ \\
\hline
Raman & $2\Gamma_{1}^+ \oplus \Gamma_{2}^+ \oplus 2\Gamma_{3}^+ \oplus \Gamma_{4}^+$  \\
Infrared & $\Gamma_{2}^- \oplus \Gamma_{3}^- $  \\
Acoustic & $\Gamma_{2}^- \oplus \Gamma_{3}^- \oplus \Gamma_{4}^-$ \\
Silent & $\Gamma_{1}^-$ \\
\hline
\hline
\end{tabular}
\end{table}
\FloatBarrier

            \item $A7$ phase [ABC stacking, $D^{5}_{3d}$ ($R\bar{3}m$, \#166) and its relation with $AB$ blue phosphorus bilayer.]
\end{enumerate}
    \end{enumerate}

 \noindent The point group for the $A7$ phase of phosphorus is $D_{3d}$, and the character table possess the same characters that the table for $AA$ blue phosphorus stacking [$D^{3}_{3d}$ ($P\bar{3}m1$, \#164)]. Due to this fact, the irreducible representations are the same and the $\Gamma^{vib}$ differs only in number of modes. The classification of the modes is given as:

\begin{table}[h!]
\caption{\label{tab:RamanIRACSilbulk} Normal vibrational mode irreducible representations ($\Gamma^{vib}$) for $A7$ phosphorus phase. Irreducible representations for the Raman active, infrared active, acoustic and silent mode are identified.}

\begin{tabular}{cc}
\hline
\hline
 \multicolumn{2}{c}{$D^{5}_{3d}$ ($R\bar{3}m$, \#166)} \\
\hline
$\Gamma^{vib}$ & $3(\Gamma_{1}^+ \oplus \Gamma_{3}^+ \oplus \Gamma_{2}^- \oplus \Gamma_{3}^-)$ \\
\hline
Raman & $3(\Gamma_{1}^+ \oplus \Gamma_{3}^+)$  \\
Infrared & $2(\Gamma_{2}^- \oplus \Gamma_{3}^-)$  \\
Acoustic & $\Gamma_{2}^- \oplus \Gamma_{3}^-$ \\
Silent & - \\
\hline
\hline
\end{tabular}
\end{table}

 \end{enumerate}

\end{document}